\documentclass[prb,aps,preprint,showpacs]{revtex4}
\usepackage{epsfig}
\usepackage{amssymb}
\usepackage{amsmath}
\usepackage{graphicx}
\usepackage{amsfonts}
\usepackage{color}
\usepackage{verbatim}

\draft
\begin{document}

\title{Anisotropy of the Spin Density Wave Onset for (TMTSF)$_2$PF$_6$ in Magnetic Field}
\author{Ya.\,A.~Gerasimenko}
\author{V.\,A.~Prudkoglyad}
\author{A.\,V.~Kornilov}
\author{V.\,M.~Pudalov}
\affiliation{P.~N.~Lebedev Physical Institute,  Moscow, 119991, Russia}
\author{V.\,N.~Zverev}
\affiliation{Institute for Solid State Physics, 142432, Chernogolovka, Moscow district, Russia}
\author{A.-K.~Klehe}
\affiliation{Clarendon Laboratory, Oxford University, OX1 3PU, UK}
\author{J.\,S.~Qualls}
\affiliation{Sonoma State University, Rohnert Park, CA 94928, USA}

\begin{abstract}
In order to study the spin density wave transition temperature ($T_{\rm SDW}$) in $\mathrm{(TMTSF)_2PF_6}$ as a function of magnetic field, we measured the magnetoresistance $R_{zz}$ in fields up to 19 T. Measurements were performed for three field orientations $\mathbf{B}\|\mathbf{a}, \mathbf{b'}$ and $\mathbf{c^*}$ at ambient pressure and at $P= 5$ kbar, that is nearly the critical pressure. For $\mathbf{B\|c^*}$ orientation we observed quadratic field dependence of $T_{\rm SDW}$ in agreement with  theory and with previous experiments. For $\mathbf{B\|b'}$ and $\mathbf{B\|a}$ orientations we have found no shift in $T_{\rm SDW}$ within 0.05 K, both at $P=0$ and $P=5$ kbar. This result is also  consistent with theoretical predictions. \end{abstract}


\date{\today}

\maketitle

\section{Introduction}
$\mathrm{(TMTSF)_2PF_6}$ is a layered organic compound that demonstrates a complex phase diagram, containing phases, characteristic of one-, two- and
three-dimensional systems. Transport properties of this material are highly anisotropic (typical ratio of the conductivity tensor components is $\sigma_{xx}:\sigma_{yy}:\sigma_{zz}\sim 10^5:10^3:1$ at $T=100$\,K \cite{Review:Lebed_Yamaji,notations}). At ambient pressure and zero magnetic field the carrier system undergoes a transition to the antiferromagnetically ordered spin density wave (SDW) state \cite{Review:Lebed_Yamaji} with a transition temperature $T_{\rm SDW}\approx12$\,K. When an external hydrostatic pressure is applied, $T_{\rm SDW}$ gradually decreases and vanishes at the critical pressure of $\sim6\,$kbar\cite{critical-pressure}. For higher pressures, $P>6$\,kbar, the SDW state is completely suppressed. Application of a sufficiently high magnetic field along the least conducting direction $\mathbf{c^*}$ restores the spin ordering. This occurs via a cascade of the field induced SDW states (FISDW) \cite{FISDW}.

\begin{figure}[ht]
\begin{center}
    \includegraphics[width=0.37\textwidth]{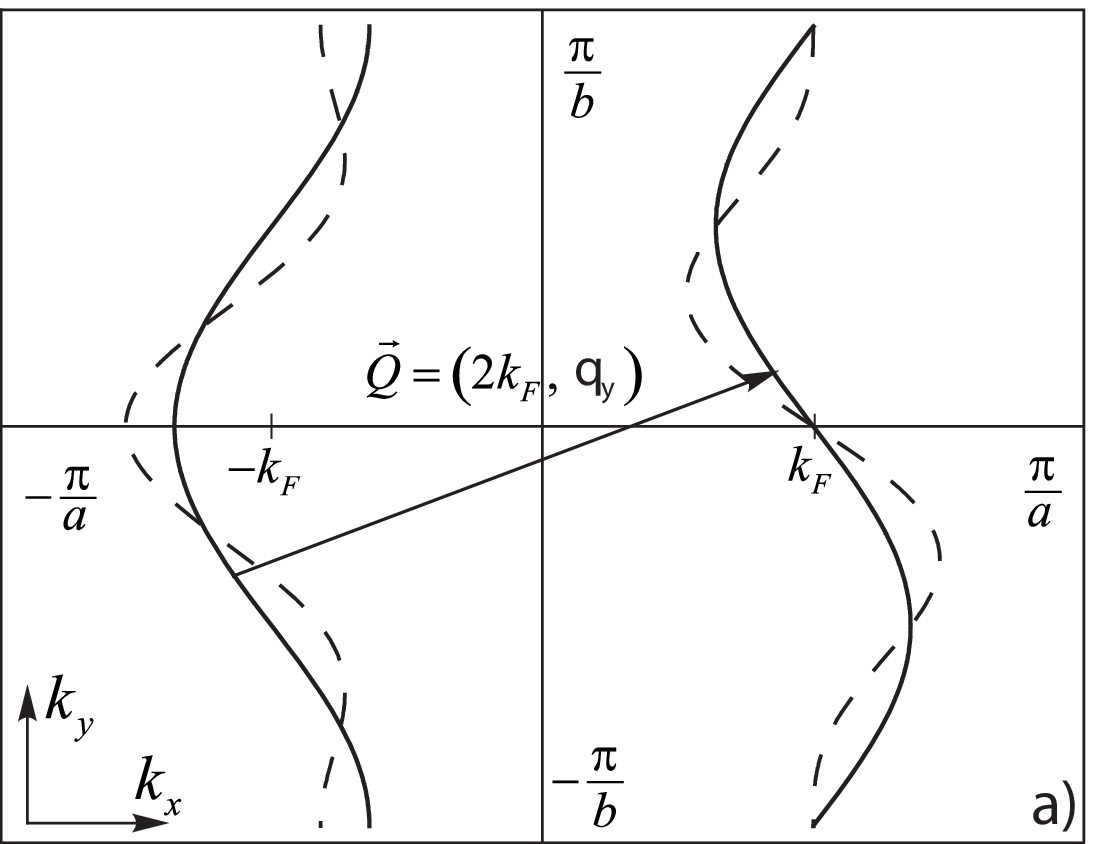}
    \includegraphics[width=0.37\textwidth]{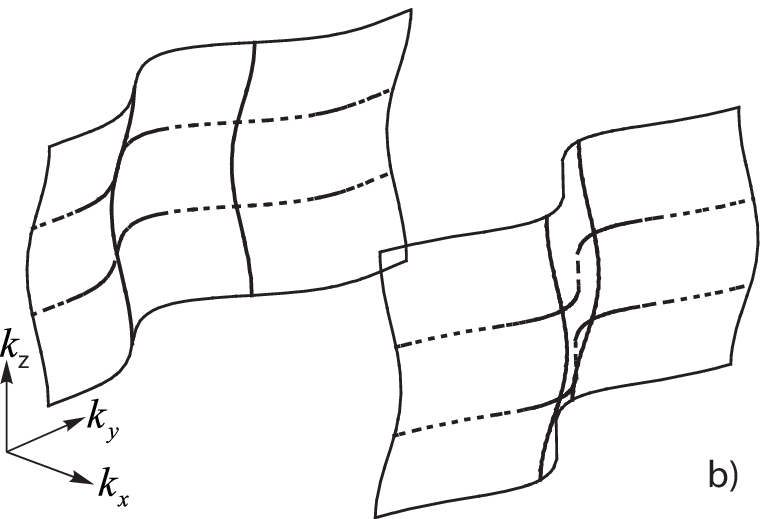}
    \caption{(a) Schematic view of the Fermi surface in the imperfect nesting model. Dashed and solid lines show FS with and without $t_b^\prime$ term in Eq.~(\ref{eqn:dlaw}) respectively ($t_b^\prime$ value is magnified for clarity). $\mathbf{Q}$  denotes the nesting vector. (b) Schematic 3D-view of the Fermi surface.Dashed and solid lines are the orbits of an electron, when magnetic field  $\mathbf{B\|c^*}$ and $\mathbf{B\|b^\prime}$ respectively.}
\label{surface}
    \end{center}
\end{figure}

The conventional model for the electronic spectrum is \cite{Gorkov_Lebed,Review:Lebed_Yamaji}:
\begin{equation}
\mathcal{E}_0(\mathbf{k})=\pm \hbar v_F(k_x \mp k_F)-2t_b \cos(k_y b')-2t_b'
\cos(2k_y b')- 2t_c \cos (k_z c^*), \label{eqn:dlaw}
\end{equation}
where $t_b,\ t_c$ are the nearest neighbor transfer integrals along $\mathbf{b^\prime}$ and $\mathbf{c^*}$ directions respectively, and $t_b'$ is the transfer integral involving next-to-nearest (second order) neighbors. For ideal one dimensional case, $t_b=t_c=t_b'=0$, and the Fermi surface consists of two parallel flat sheets. This surface satisfies the so-called ideal nesting condition: there exists a vector $\mathbf{Q}_0$ which couples all states across the Fermi surface. In the quasi-one dimensional case, when $t_b$ and $t_c$ are non-zero, the Fermi-sheets become slightly corrugated. Nevertheless, one can still find a vector, that couples all states across the Fermi surface, therefore the ideal nesting property also holds in this case. It means that the magnetic susceptibility $\chi(\mathbf{q})$ of the system diverges at $\mathbf{q}=\mathbf{Q}_0$ and the system is unstable against formation of SDW \cite{Review:Lebed_Yamaji}. When $t_b'$ is non-zero, the situation changes drastically: no vector can couple all states on both sides of the Fermi surface, though $\mathbf{Q}_0$ still couples a large number of states. The situation called ``imperfect nesting'' is sketched on Fig.~\ref{surface}a.

Despite the complex behavior of the system, theory\cite{Montambaux, Maki} successfully describes the effects of pressure and magnetic field on the SDW
transition in terms of the single parameter $t_b'$. According to the theory, $t_b'$ increases with external pressure, and conditions for nesting deteriorate. Therefore, under pressure deviations of the system from the ideal 1D-model become more prominent and, as a consequence, $T_{\rm SDW}$ decreases. When $t_b'$ reaches a critical value $t_b^*$, the SDW transition vanishes. The application of a magnetic field normal to the $\mathbf{a}$ direction restricts electron motion in the $\mathbf{b\mathrm{-}c}$ plane making the system effectively more one dimensional. Theory Ref.~\onlinecite{Montambaux} predicts that the transition temperature increases in weak  fields $\mathbf{B\|c^*}$ as
\begin{equation}
\nonumber \Delta T_{\rm SDW}(B)=T_{\mathrm{SDW}}(B)-T_{\mathrm{SDW}}(0)=\alpha B^2,
\end{equation}
and further saturates in high fields; here $\alpha=\alpha(P)$ is  a function of pressure.

A number of experiments\cite{critical-pressure,Chaikin_abc,Tsdw-Biskup,Tsdw-highfields} were made to examine the predictions of the theory for the $\mathbf{B\|c^*}$ case. All these studies confirmed quadratic field dependence of the transition temperature. Nevertheless, the predicted saturation has not been seen until now. Furthermore, Murata et. al \cite{uniaxial-1,uniaxial-2,uniaxial-3,uniaxial-4} reported an unexpected anisotropy of the $T_{\rm SDW}$ in $\mathrm{(TMTSF)_2PF_6}$ under uniaxial stress, the result seems to disagree with the theory. According to theory \cite{Montambaux,Maki}, the only relevant parameter is $t^\prime_b$; therefore, one might expect the uniaxial stress along $\mathbf{b}^\prime$ to affect $T_{\rm SDW}$ stronger than the stress in other directions. Murata et al.\cite{uniaxial-1,uniaxial-2,uniaxial-3,uniaxial-4}, however, showed experimentally that the uniaxial stress applied along the $\mathbf{a}$ direction changed $T_{\rm SDW}$ stronger than the stress in the $\mathbf{b}^\prime$ direction.

The results mentioned above demonstrate that the consistency between the theoretical description and experiment is incomplete. Whereas there is a number of experimental data for the magnetic field $\mathbf{B\|c^*}$, for $\mathbf{B\|a}$ and $\mathbf{B\|b'}$ only one experiment\cite{Chaikin_abc}
has been done so far at ambient pressure, and none at elevated pressure. Danner et al.\cite{Chaikin_abc} observed no field dependence
for $\mathbf{B\|a}$ and $\mathbf{B\|b'}$ at ambient pressure. The absence of a field dependence, however, cannot be considered as a crucial test of the theory, because the effect of the magnetic field might be small at ambient pressure. Indeed, according to the theory, elevated pressure enhances any imperfections of nesting, and the effect of magnetic field is expected to become stronger. As a result, the strongest effect should take place at pressures close to the critical value. The aim of the present work, therefore, is to determine experimentally $T_{\rm SDW}(B)$ dependence for $\mathbf{B\|a}$ and $\mathbf{B\|b'}$ near the critical pressure.

We report here our measurements of the magnetic field induced shift in $T_{\rm SDW}$ made at $P=0$ and 5\,kbar for the three orientations $\mathbf{B\|a, B\|b',\mbox{ and } B\|c^*}$. Our main result is that for $\mathbf{B\|a}$ and $\mathbf{B\|b^\prime}$ there is no distinct shift of the transition temperature within our measurements' uncertainty 0.05\,K at pressure up to 5\,kbar and in fields up to 19\,T. At the same time, we found quadratic $T_{\rm SDW}(B)$ dependences for $\mathbf{B\|c^*}$ both at zero and non-zero pressures, a result in agreement with previous studies by other groups \cite{critical-pressure,Chaikin_abc,Tsdw-Biskup,Tsdw-highfields}. We suggest an explanation of our experimental data, based on the mean-field
theory, and show that the latter correctly describes the effect of the magnetic field on $T_{\rm SDW}$.

\section{Experimental}
Single-crystal samples of $\mathrm{(TMTSF)_2PF_6}$ were grown by conventional electrochemical technique. Measurements were made on three samples from the same batch (the typical sample size is $3\times0.25\times0.1\ \mathrm{mm}^3$ along $\mathbf{a},\mathbf{b^\prime}$ and $\mathbf{c^*}$ directions respectively). Eight fine wires (10\,$\mu m$ Au wires or 25\,$\mu m$ Pt wires) were attached to the sample with conductive graphite paste. Two groups of four contacts were made on the two opposite a-b$^\prime$ faces of the sample along a-axis. All measurements were made by four-probe ac lock-in technique at 10-120\,Hz frequencies. The out-of-phase component of the contacts resistance was negligible. The resistance along c$^*$-axis, $R_{zz}$, was measured using two pairs of contacts on top and bottom faces, normal to the c$^*$-axis.

For measurements under pressure the sample and a manganin pressure gauge were inserted into a miniature nonmagnetic spherical pressure cell\cite{PressureCell} with an outer diameter of 15\,mm. The cell was filled with Si-organic pressure transmitting liquid\cite{Si-organic_liquid} (PES-1). The pressure was applied and fixed at room temperature. The pressure values quoted throughout this paper refer to those determined at helium temperature. After pressure was applied, the cell was mounted in a two-axis rotation stage placed in liquid $^4$He in a bore of a 21\,T superconducting magnet at Oxford University. The rotating system enabled rotation of the pressure cell around the main axis by $200^\circ$ (with uncertainty of $\sim 0.1^\circ$) and around the auxiliary axis by $360^\circ$ (with uncertainty of $\sim 1^\circ$); this allowed us to set the sample at any desired orientation with respect to the magnetic field direction.

Measurements at ambient pressure were performed using more simple rotating
system which allowed rotation around only one axis (perpendicular to the field
direction) by $\sim200^\circ$ with $\sim 0.1^\circ$ uncertainty. This system
was mounted in a bore of a 17\,T superconducting magnet at ISSP.

\begin{figure}[ht]
\begin{center}
    \includegraphics[width=0.47\textwidth]{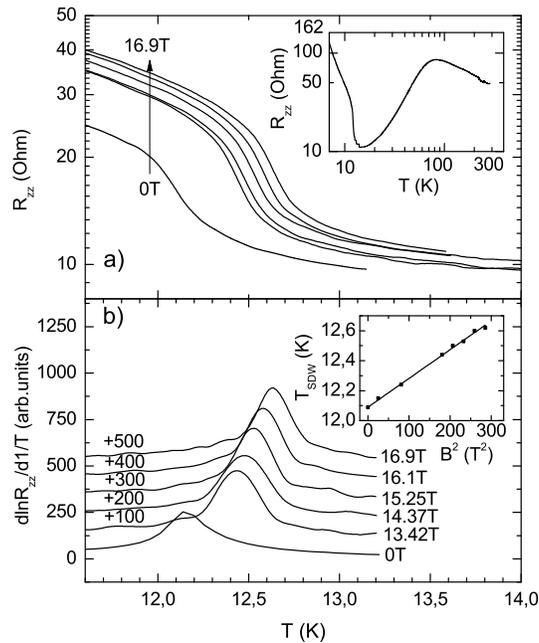}
    \caption{Temperature dependences of $R_{zz}$ at ambient pressure for six values of magnetic field (as shown on panel (b)) aligned with the least conduction direction, $\mathbf{B\|c^*}$. (a) $R_{zz}(T)$ for the set of magnetic fields; (b) logarithmic derivatives of the same data. Inset to panel (a) demonstrates typical dependence of $R_{zz}$ versus T. Inset to (b) shows linear fit for transition temperatures, obtained from the derivative plots, vs $B^2$.}
\label{ambient}
    \end{center}
\end{figure}

\begin{figure}[ht]
\begin{center}
    \includegraphics[width=0.47\textwidth]{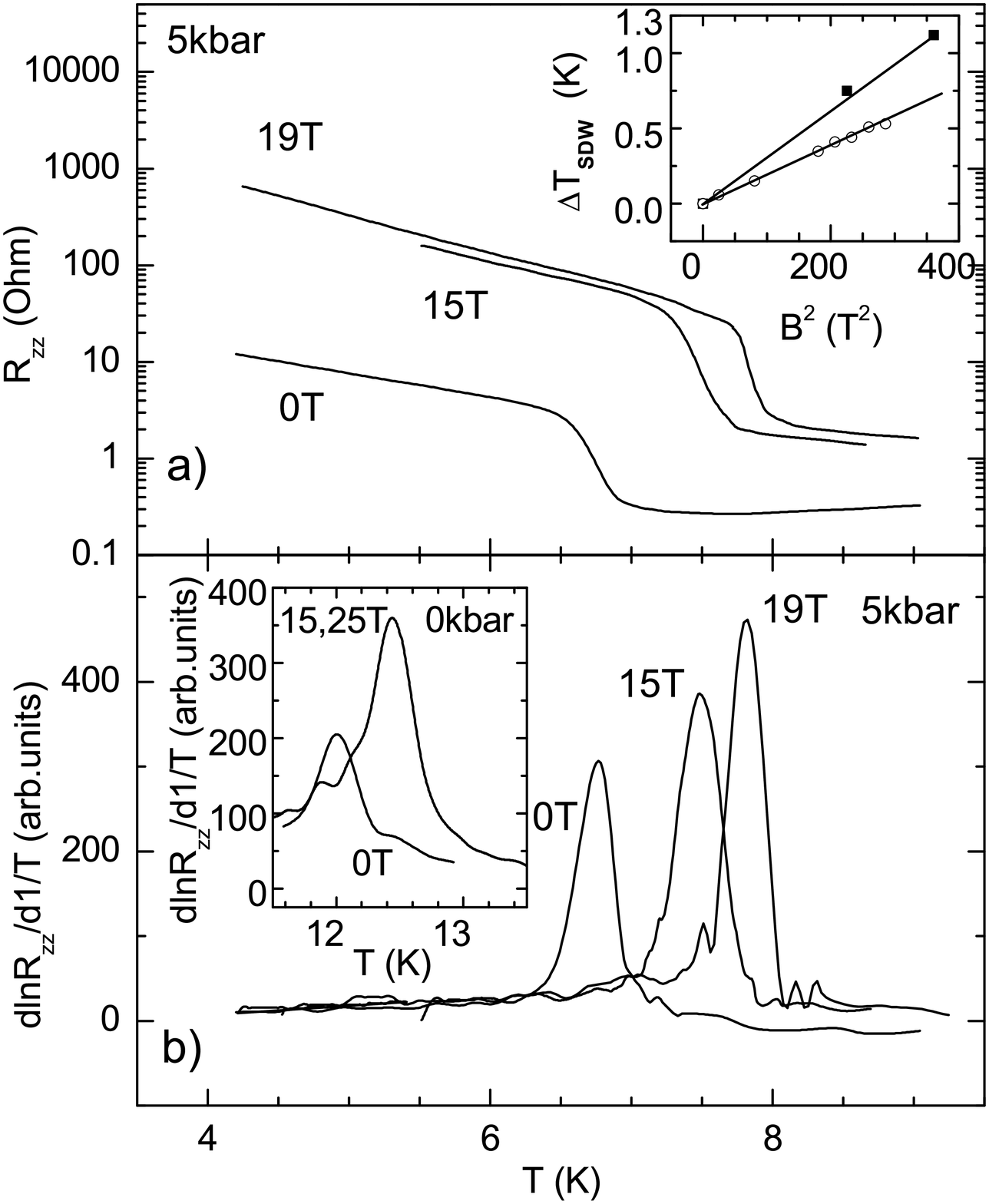}
    \caption{Temperature dependences of $R_{zz}$ at an elevated pressure of 5\,kbar for a field orientation $\mathbf{B\|c^*}$. (a) $R_{zz}(T)$ for the set of magnetic fields. The inset shows  $\Delta T_{SDW}$ versus $B^2$ for $P=5\,$kbar (filled dots) and for $P=0$ (empty circles). (b) logarithmic derivatives of the same data. The inset demonstrates that $T_{SDW}$ shift in the magnetic field  is much more pronounced at $P=5$\,kbar than that at ambient pressure.}
\label{pressure}
    \end{center}
\end{figure}

\begin{figure}[ht]
\begin{center}
    \includegraphics[width=0.47\textwidth]{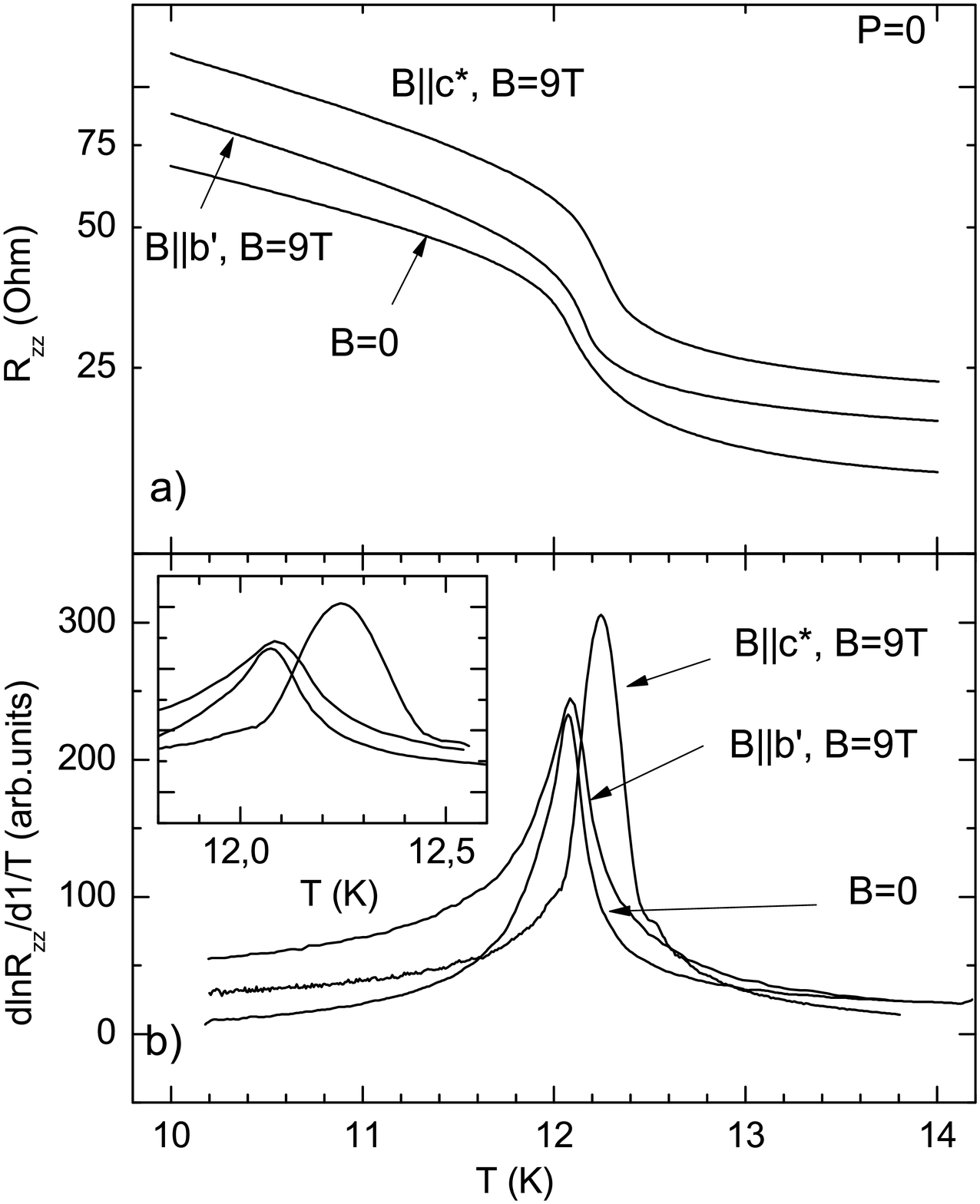}
    \caption{Temperature dependences of $R_{zz}$ at ambient pressure in magnetic field 9\,T compared for $\mathbf{B\|c^*}$ and $\mathbf{B\|b'}$ orientations. (a) panel shows $R_{zz}(T)$, (b) panel shows logarithmic derivatives of these dependences. Derivative graphs in a larger scale are shown on the inset.}
\label{b-ambient}
    \end{center}
\end{figure}

\begin{figure}[ht]
\begin{center}
    \includegraphics[width=0.45\textwidth]{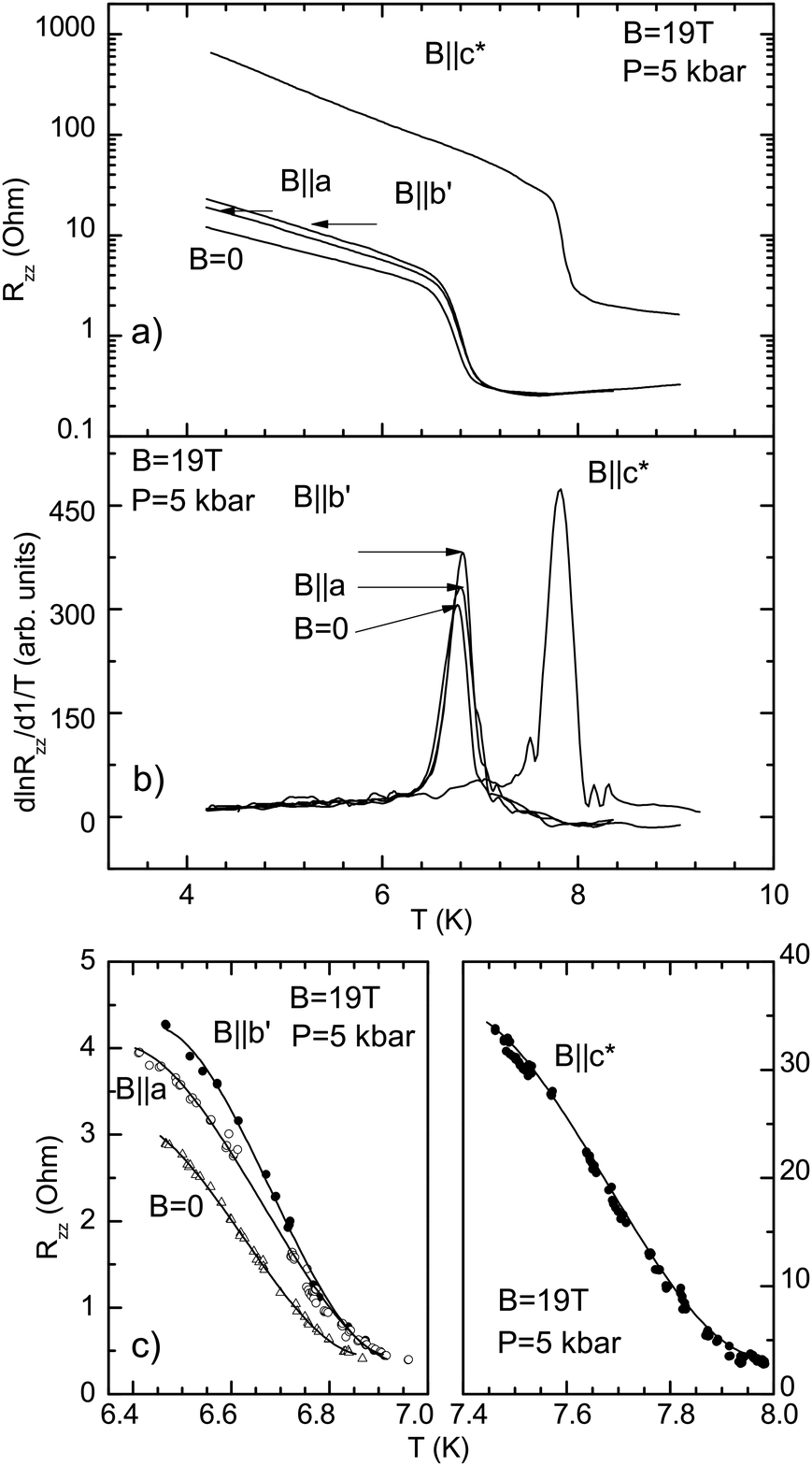}
    \caption{Temperature dependences of $R_{zz}$ under pressure $P=5$\,kbar in magnetic field 19\,T aligned with $\mathbf{B\|a}$ or $\mathbf{B\|b'}$ compared with orienation $\mathbf{B\|c^*}$; (a) and (b) panels show $R_{zz}(T)$ and their logarithmic derivatives respectively. These results are corrected due to magnetoresistance of RuO$_2$ thermometer. Panel (c) zooms in 0.5\,K interval near the transition, solid lines are cubic polynomial fits of experimental points.}
\label{orientation}
    \end{center}
\end{figure}
Samples were cooled very slowly, at the rate of $0.2\div0.3$\,K/min to avoid microcracks. Nevertheless, some samples cooled down at ambient pressure
experienced 1-2 microcracks, seen as an irreversible jumps (a few percents) in the sample resistance. No cracks were observed during cooling of a sample in the pressure cell. During measurements under pressure the temperature of the cell was determined by RuO$_2$ thermometer, and during measurements at ambient pressure -- by Cu-Fe-Cu thermocouple and RuO$_2$ thermometer. The temperature was varied slowly in order to insure, that the sample and the thermometer were in thermal equilibrium. The thermal equilibrium condition was verified by the absence of a hysteresis in $R_{zz}(T)$ between cooling and heating cycles.
\section{Results and discussion}
Measurements were performed on three samples from the same batch and the results were in qualitative correspondence with each other. Most detailed data taken for two samples are presented in this section.
\subsection{$\mathbf{B\|c^*}$}
Fig.~\ref{ambient} shows the temperature dependence of $R_{zz}$ at ambient pressure and different magnetic fields. $R_{zz}(B=0)$ at ambient pressure is shown in the inset to Fig.~\ref{ambient}a over a large temperature range: as temperature decreases, the resistance decreases monotonically, then exhibits a sharp jump and further increases in a temperature activated manner. The jump at 12\,K indicates the transition to the
low-temperature spin-density wave state. Throughout the paper we define the transition temperature according to the peak in $d\ln R_{zz}/d(1/T)$, the logarithmic derivative of resistance \emph{vs} inverse temperature. As the magnetic field applied along $\mathbf{c^*}$-axis  grows, $T_{\rm SDW}$ is shifted progressively to higher temperatures (Fig.~\ref{ambient}b). The shift increases quadratically with field, as shown in the inset to Fig.~\ref{ambient}b.

Application of pressure $P=5\,$kbar lowers the zero-field transition temperature down to 6.75\,K (Fig.~\ref{pressure}a,~b). The pressure dependence of $T_{\rm SDW}(P)$ is known to be strongly nonlinear \cite{chaikin93}, its slope is small at low pressures and sharply increases in the vicinity of the critical pressure value. Therefore, the factor of two decrease in $T_{\rm SDW}(P)$ (from 12\,K to 6.75\,K, compare Figs. 2 and 3) demonstrates  that the pressure is close to the critical value. At a pressure of 5\,kbar and in the presence of a magnetic field $\mathbf{B\|c^*}$, the transition temperature $T_{\rm SDW}$ grows nearly quadratically with field, $\Delta T_{\rm SDW}\propto B^2$  (see inset to Fig.~\ref{pressure}a). This growth is qualitatively similar to that for zero pressure, however, it is much more pronounced, compared to the former case (see Fig.~\ref{pressure}b and the inset to Fig.~\ref{pressure}a).

Application of a magnetic field also increases resistance in the SDW state (cf. Figs.~\ref{ambient}a,\,\ref{pressure}a). In principle, the resistance growth should be related with the increase of $T_{\rm SDW}$, e.g. due to the increase of the SDW gap in magnetic field\cite{Maki}. However, the data on Fig.~\ref{pressure} as well as previous observations (for example, Ref.~\onlinecite{Tsdw-highfields}) indicate, that $R_{zz}(T)$ cannot be described by a temperature-activated behavior, both in zero and non-zero magnetic fields. Apparently, the observed $R_{zz}(T,B)$ dependence is governed by both the increase of the SDW intensity in magnetic field and the magnetoresistance. Therefore, without an adequate model for $R_{zz}(T,B)$ the two contributions cannot be separated, even though our data clearly indicate the correlation of the resistance growth and the increase of $T_{\rm SDW}$ in magnetic field.

The observed $T_{\rm SDW}(B)$ dependence (Fig.~\ref{pressure}) for our samples in magnetic field along $\mathbf{c^*}$ is qualitatively consistent with
theory\cite{Montambaux,Maki} and with earlier observations by other groups \cite{critical-pressure,Chaikin_abc,Tsdw-Biskup,Tsdw-highfields}.
According to the theory, pressure deteriorates nesting conditions, enhancing the $t_b^\prime$ term in the energy spectrum Eq.~(1). Therefore, under pressure the number of unnested electrons increases as compared to the ambient pressure case. In contrast to the action of pressure, application of magnetic field $\mathbf{B\|c^*}$ improves the nesting conditions, both at elevated and at zero pressure, although the number of unnested electrons is larger in the former case. This is predicted to lead to an enhancement of the field dependence of $T_{\rm SDW}$ under pressure.

Our data (see inset to Fig.~\ref{pressure}a) confirms the theoretically predicted enhancement of the $T_{\rm SDW}(\mathbf{B\|c^*})$ dependence at pressures close the critical value. We therefore anticipate, that if the  $T_{\rm SDW}(B)$ dependence existed for other field orientations, it would be enhanced at elevated pressures. Correspondingly, we performed an experimental search  for this dependence at $P$ close to the critical value of $P_c$.
\subsection{$\mathbf{B\|a,b^\prime}$}
When the magnetic field is applied in the a-b plane, it's effect on the SDW-transition temperature is  either  missing or, at least, is  much less than for $\mathbf{B\|c^*}$. Figures 4\,a,\,b illustrate this result for one of the orientations, $\mathbf{B\|b^\prime}$. Even though the shape of the $R_{zz}(T)$  curves slightly changes   with field, the temperature of the transition remains unchanged within our measurement uncertainty of $\sim0.03\,$K. For comparison, on the same figures we present also the $R_{zz}(T)$ data for  the $\mathbf{B\|c^*}$ orientation, demonstrating that the shift of $T_{\rm SDW}$ in the same field of 9\,T is an order of magnitude higher for $\mathbf{B\|c^*}$.

In line with the experimental situation for the $\mathbf{B\|c^*}$, one might expect the shift in $T_{\rm SDW}$ (if any) to be enhanced under pressure. Figures~\ref{orientation}a,b summarize the  main result of our paper --- the $R_{zz}(T)$  dependences across the transition measured for all three field orientations ($\mathbf{B\|a}$, $\mathbf{b^\prime}$, and $\mathbf{c^*}$) at $P= 5$\,kbar, close to the critical pressure. At $P=5$\,kbar and at $B=19$\,T, the shift $\Delta T_{\rm SDW} (B)$ is as large as 1\,K for $\mathbf{B\|c^*}$, whereas for $\mathbf{B\|a,b^\prime}$ the shift is either missing or vanishingly small , at least a factor of 20 smaller than for $\mathbf{B\|c^*}$ (see Fig.~5\,b).

Zooming the data in Fig.~5\,c (on the left panel), one can notice that $R_{zz}(T)$ curves for $\mathbf{B\|a,b^\prime}$ are slightly shifted from the $\mathbf{B}=0$ one. However, our measurements uncertainty is comparable with this difference; for this reason, the sources of this uncertainty are analyzed below.

There are two possible sources of uncertainties: (i) the calibration error of the RuO$_2$ thermometer in magnetic fields, and (ii) an uncertainty of the procedure used to determine the transition temperature. The latter contribution was determined by the width of the transition and was estimated to be about 0.02 - 0.03\,K.  As for the former one, in all measurements we used RuO$_2$ resistance thermometer whose magnetoresistance was calibrated at 4.2\,K. Possible changes of the RuO$_2$ magnetoresistance between 4.2\,K and  6.7\,K are the major source of our uncertainty and are estimated to be 0.04\,K. Thus, we can only quantify the changes $\Delta T_{\rm SDW}(B)$ that are larger than 50\,mK. If the transition temperature changes with $\mathbf{B\|a}$ or $\mathbf{B\|b^\prime}$, the changes are to be smaller than the above value.
\subsection{Discussion}
(i) In theory \cite{Montambaux,Maki}, the changes of the transition temperature result from imperfect nesting. The energy spectrum Eq.~(1) contains the only term $t_b^\prime\cos(k_yb^\prime)$  that is responsible for the nesting imperfection. Magnetic field parallel to the $\mathbf{c^*}$ direction eliminates the electron dispersion in the $\mathbf{b^\prime}$ direction from the system Hamiltonian \cite{Chaikin_q1d}. This effect is somewhat similar to the quasiclassical action of the Lorentz force on the electrons. The force is directed along $\mathbf{b^\prime}$ axis because the Fermi velocity $\mathbf{v}_F$ in (TMTSF)$_2$PF$_6$ is along the $\mathbf{a}$-axis, on average. It makes electrons on the Fermi surface cross the Brillouin zone in the $k_y$ direction (see Fig.~\ref{surface}a). Such a motion averages the electron's energy over all $k_y$ states \cite{Chaikin_q1d}, and all the terms, that contain $\cos(k_yb^\prime)$ in the electron spectrum Eq.~(\ref{eqn:dlaw}), vanish. Since the $t_b^\prime$ term, responsible for the nesting imperfection, also vanishes, the magnetic field $\mathbf{B\|c^*}$ improves nesting conditions; this results in a growth of the transition temperature. This effect is described by the mean-field theory \cite{Montambaux,Maki}. In contrast, a magnetic field $\mathbf{B\|b^\prime}$ has no effect on $t_b^\prime$, therefore, no shift in $T_{\rm SDW}$ should occur in this field orientation. Our result that for $\mathbf{B\|b^\prime}$ the shift in $T_{\rm SDW}$ is much less than for $\mathbf{B\|c^*}$ does not contradict this prediction.

(ii) In principle, the magnetic field $\mathbf{B\|b^\prime}$ still can affect the electron dispersion in the $\mathbf{c^*}$ direction. In theory \cite{Montambaux} such a dispersion is neglected and $t_b^\prime$ is assumed to be the only term responsible for imperfect nesting. In general, besides $t_b^\prime$ there are other antinesting terms, that can affect $T_{\rm SDW}$ in field $\mathbf{B\|b^\prime}$. Studies of $T_{\rm SDW}$ anisotropy for different field direction may in principle provide information on the $t_b/t_c$ ratio. In what follows, we estimate the $t_b/t_c$ ratio from our experimental data. In order to do this, we expand the energy spectrum Eq.~(\ref{eqn:dlaw}):
\begin{equation}
\label{eqn:dlaw_ext}
\mathcal{E}_1(\mathbf{k})=\mathcal{E}_0(\mathbf{k})- 2t_{bc}^\prime \cos(k_y b^\prime)\cos(k_z c^*)-2t_c^\prime \cos(2k_zc^*),
\end{equation}
where $t_{bc}^\prime$ and $t_c^\prime$ are the next-to-nearest hopping integrals. In the model Eq.~(\ref{eqn:dlaw_ext}) and for magnetic field direction $\mathbf{B\|c^*}$, the correction to $\Delta T_{\rm SDW}$ from the $t_{bc}^\prime$ term is considerably smaller, than that from $t_b^\prime$, for $t_b/t_c\gg1$. Therefore, the $T_{\rm SDW}(\mathbf{B\|c^*})$ dependence is almost unchanged, when $t_{bc}^\prime$ is taken into account.

However, for $\mathbf{B\|b^\prime}$ the situation is essentially different. The electrons experience now the Lorenz force along $\mathbf{c^*}$ axis, and the corresponding motion along $k_z$  averages out all $\cos(k_z c^*)$ terms in the electron spectrum Eq.~(\ref{eqn:dlaw_ext}). Therefore, the contribution of $t_{bc}^\prime$ to $T_{\rm SDW}(B)$ dependence becomes dominant.

(iii) When magnetic field is applied along $\mathbf{a}$-axis, it is not expected to alter the electron motion, because the Lorentz force is zero, on average. Correspondingly, there are no terms in the electron spectrum which may be affected by the magnetic field in this orientation  and the transition temperature is not expected to depend on the field $\mathbf{B\|a}$.

From the above discussion we conclude that $T_{\rm SDW}$ in principle might be affected by the field $\mathbf{B\|b^\prime}$. Bjeli$\mathrm{\check s}$ and Maki in Ref.~\onlinecite{Bjelis_Maki} took the $t_{bc}^\prime$ term into account and derived an expression for the transition temperature in tilted magnetic field. Based on this result, one can show (see Appendix) that in high magnetic fields the anisotropy of $\Delta T_{\rm SDW}(\mathbf{B})$ is related with the $t_b/t_c$ ratio:
\begin{equation}\label{tdif2}
    \frac{T_{\rm SDW}(B\|c^*)-T_{\rm SDW}(0)}{T_{\rm SDW}(B\|b^\prime)-T_{\rm SDW}(0)}\approx\beta\frac{1}{4}\left(\frac{t_b}{t_c}\right)^2\left(\frac{\omega_c}{\omega_b}\right)^2,
\end{equation}
where $\beta\approx1$ is a numeric factor. Since the above relationship is dominated by $(t_b/t_c)^2$, the shift in $T_{\rm SDW}$ for $\mathbf{B\|b^\prime}$ is expected to be considerably weaker than for $\mathbf{B\|c^*}$.

Fig.~\ref{orientation} shows the experimental data for SDW transition in fields $\mathbf{B}\|\mathbf{a},\,\mathbf{b^\prime},\,\mathbf{c^*}$. This data enables us to estimate the $t_b/t_c$ ratio using Eq.~(\ref{tdif2})\cite{scattering}. However, such a straightforward comparison of $T_{\rm SDW}$ in $B=19$\,T and $B=0$ includes a large uncertainty related with thermometer magnetoresistance. In order to overcome this problem, we ramped the temperature slowly in a fixed magnetic field 19\,T and measured $R_{zz}(T)$. We repeated the procedure for the three field orientations ($\mathbf{B}\|\mathbf{a},\,\mathbf{b^\prime},\,\mathbf{c^*}$) by rotating in situ the pressure cell with the sample with respect to magnetic field direction. The $T_{\rm SDW}(\mathbf{B})$ data measured this way were used to calculate $t_b/t_c$ with Eq.~(\ref{tdif2}). In the calculations we substituted $T_{\rm SDW}(B\|a)$ for $T_{\rm SDW}(0)$ because as discussed in (iii) magnetic field $\mathbf{B\|a}$ does not affect $T_{\rm SDW}$. Such a procedure enabled us to eliminate the error related with the magnetoresistance of RuO$_2$ thermometer. Yet the difference $\Delta T_{ab}=T_{\rm SDW}({B\|b^\prime})- T_{\rm SDW}({B\|a})$ was within the error bar of $0.02$\,K in the experiment, the upper bound of our estimate $\Delta T_{ab}=0.02$\,K corresponds according to Eq.~(\ref{tdif2}) to the lower bound of $t_b/t_c\approx7$. This estimate agrees with earlier result $t_b/t_c\approx6$ obtained from angle-dependent magnetoresistance studies in the metallic state at 7\,kbar\cite{Naughton}. The $t_b/t_c$ estimates indicate that the contribution of the $t_{bc}^\prime$ term to $T_{\rm SDW}$ is negligible, a factor of 50 smaller than that of $t_b^\prime$.

\section{Conclusion.}
In conclusion, we have measured the magnetic field effect on the transition temperature $T_{\rm SDW}$ to the spin density wave state in (TMTSF)$_2$PF$_6$ in fields up to $\mathbf{B}=19$\,T for three orientations $\mathbf{B\|a, b', c^*}$, and at pressures up to 5\,kbar. Measurements for $\mathbf{B\|c^*}$ are in qualitative agreement with the mean field theory \cite{Montambaux, Maki} and with results of other groups\cite{critical-pressure,Chaikin_abc,Tsdw-Biskup,Tsdw-highfields}. Our data confirm that the field dependence of $T_{\rm SDW}$ is enhanced as
pressure increases and approaches the critical value. Measurements of $T_{\rm SDW}$ for $\mathbf{B\|a,b^\prime}$ under pressure are presented for the first time. The main result of our paper is that the magnetic field dependence of $T_{\rm SDW}$ for $\mathbf{B\|a}$ and for $\mathbf{B\|b^\prime}$ is either absent or vanishingly small (at least a factor of 20 smaller than for $\mathbf{B\|c^*}$) even near the critical pressure and at $B=19$\,T. This shows, that the influence of other imperfect nesting terms on $T_{\rm SDW}$ is negligibly small. This result confirms the assumption of the theory that $T_{\rm SDW}$ is determined by the antinesting terms with the biggest contribution from the $t^\prime_b$ term in the electron spectrum.
\section{Acknowledgements}
We are grateful to P.\,D.~Grigoriev, A.\,G.~Lebed and A.~Ardavan for their valuable suggestions and discussion of our results. The work was partially supported by the Programs of the Russian Academy of Sciences, RFBR (08-02-01047, 09-02-12206), Russian Ministry of Education and Science, the State Program of support of leading scientific schools (1160.2008.2), EPSRC, and the Royal Society.

\section{Appendix: derivation of Eq.~(\ref{tdif2})}
Bjeli$\mathrm{\check s}$ and Maki in Ref.~\onlinecite{Bjelis_Maki} took the $t_{bc}^\prime$ term into account and derived an expression for the transition temperature for magnetic field $\mathbf{B}$ in $\mathbf{b^\prime}-\mathbf{c^*}$ plane. General form of this expression involves series of products of Bessel functions. In high magnetic fields the expression can be simplified by saving only the greatest term in the series. Namely, for $\mathbf{B\|c^*}$:
\begin{equation}
\ln\left[\frac{T_\mathrm{SDW}(B\|c^*)}{T_\mathrm{SDW}}\right]\approx J_1^2\left(\frac{t_{b}^\prime}{\omega_b}\right)J_0^2\left(\frac{t_{bc}^\prime}{\omega_b}\right)\left[\mathrm{Re}\Psi\left(\frac{1}{2}+\frac{i2\omega_b}{4\pi T_\mathrm{SDW}}\right)-\Psi\left(\frac{1}{2}\right)\right]\label{eqn:tcc}
\end{equation}
and for $\mathbf{B\|b^\prime}$:
\begin{equation}
\ln\left[\frac{T_\mathrm{SDW}(B\|b^\prime)}{T_\mathrm{SDW}}\right]\approx J_1^2\left(\frac{t_{bc}^\prime}{\omega_c}\right)\left[\mathrm{Re}\Psi\left(\frac{1}{2}+\frac{i\omega_c}{4\pi T_\mathrm{SDW}}\right)-\Psi\left(\frac{1}{2}\right)\right]\label{eqn:tcb}
\end{equation}
Here $T_\mathrm{SDW}=T_\mathrm{SDW}(B=0)$, $J_{0,1}$ are Bessel functions and $\Psi$ is a digamma function. When $\mathbf{B\|c^*}$, Lorentz force pushes the electrons to cross the Brillouin zone in $k_y$ direction with the characteristic frequency of $\omega_b=ev_FBb$ (see Fig.\ref{surface}b). When $\mathbf{B\|b^\prime}$ the frequency is $\omega_c=ev_FBc$, a factor of 2 larger than $\omega_b$. Substitution of the lattice parameters and $v_F\sim1.1\cdot10^5$\,[m/sec]\cite{Tsdw-highfields} gives $\omega_b\approx0.985\cdot B$\,[K]. Therefore, in high fields $t_b^\prime/\omega_b$ and $t_{bc}^\prime/\omega_b$ vanish, leaving only terms with lower order Bessel functions in the original series. Consequently, we arrive at Eq.~(\ref{eqn:tcc}) and (\ref{eqn:tcb}).

One can show by expanding exponents in series, that
\begin{equation}\label{tdif}
    \frac{T_{\rm SDW}(B\|c^*)-T_{\rm SDW}(0)}{T_{\rm SDW}(B\|b^\prime)-T_{\rm SDW}(0)}\approx\beta\frac{J_1^2\left(\frac{t_b^\prime}{\omega_b}\right)J_0^2\left(\frac{t_{bc}^\prime}{\omega_b}\right)}{J_1^2\left(\frac{t_{bc}^\prime}{\omega_c}\right)},
\end{equation}
where $\beta\approx1$ is a numeric factor. By substituting asymptotic forms for Bessel functions with small arguments, one can obtain
\begin{equation}\label{tdif22}
    \frac{T_{\rm SDW}(B\|c^*)-T_{\rm SDW}(0)}{T_{\rm SDW}(B\|b^\prime)-T_{\rm SDW}(0)}\approx\beta\frac{1}{4}\left(\frac{t_b}{t_c}\right)^2\left(\frac{\omega_c}{\omega_b}\right)^2,
\end{equation}
Therefore, by measuring the above differences of the transition temperature one can determine the ratio of the transfer integrals $t_b/t_c$.


\begin{thebibliography}{99}
\bibitem{notations}Through this paper we use the approxiamtion of orthorhombic elementary cell with the basic vectors $\mathbf{a}$, $\mathbf{b^\prime}$ and $\mathbf{c^*}$ and the coordinates $(x,y,z)$ in corresponding directions.
\bibitem{scattering}Strictly speaking, theoretical models\protect\cite{Montambaux,Maki} for $T_{\rm SDW}(B)$ do not take finite scattering time $\tau$ into account; however we expect Eq.~\protect\ref{tdif2} to be valid in the high-field limit, $\omega_c\tau\gg1$. For the studied sample we observed the FISDW transitions starting from field $B=7$\,T (for $T=2$\,K and $P=10$\,kbar). This result provides the evidence that $\omega_c\tau\gg1$ for our data taken at 19\,T and thus ensures the applicability of Eq.~\protect\ref{tdif2}.
\end{thebibliography}
\end{document}